\documentclass[aps]{revtex4}
\usepackage{graphicx}
\begin{document}

\title{A Note on Transverse Axial Vector and Vector Anomalies in
$U(1)$ Gauge Theories}
\author{Wei-Min Sun$^1$, Hong-Shi Zong$^{2,1}$,
Xiang-Song Chen$^1$ and Fan Wang$^1$}
\affiliation{$^1$Department of Physics and
Center for Theoretical Physics,
Nanjing University, Nanjing 210093, China\\
$^2$CCAST(World Laboratory), P.O. Box 8730, Beijing 100080, China}

\begin{abstract}
The transverse axial vector and vector 
anomalies in four-dimensional $U(1)$ gauge theories studied in 
\cite{HeAnomaly} is reexamined by means of perturbative method. The 
absence of transverse
anomalies for both axial vector and vector current is verified.
We also show that the Pauli-Villars regularization and dimensional
regularization gives the same result on the transverse anomaly of both 
axial vector and vector curent.
\end{abstract}

\maketitle

Whenever a quantum field theory possesses some symmetry, the Green functions 
of the theory obey a series of exact relations generally known as 
Ward-Takahashi(WT) identities \cite{WardTakahashi}. They play an important
role in various problems in the study of quantum field theory, such as
the proof of renormalizability of gauge theories \cite{Itzykson}.
In the Dyson-Schwinger equation(DSE) approach to gauge theories 
\cite{Roberts,Alkofer}
 the fermion-boson
vertex function is a very difficult quantity to be specified. Here the WT identities
are used to constrain
the form of the vertex function. However, the 
normal
WT identity only specifies the longitudinal part of the vertex 
function, leaving
its transverse part undetermined. More specifically, the longitudinal part of the 
fermion-boson vertex function can be written in terms of the full fermion 
propagator
as 
\begin{equation}
k_\mu \Gamma^{\mu}(q,p)=S^{-1}(q)-S^{-1}(p),~~~k_\mu \equiv q_\mu-p_\mu
\end{equation}
It is evident that the above WT identity does not at all specify the transverse 
part of the vertex.
In order to obtain further constraints on the vertex function, some authors 
have studied the
so-called transverse WT identities \cite{Kondo,He} which specify the curl 
of the vertex
$\partial_\mu \Gamma_\nu-\partial_\nu \Gamma_\mu$.

However, the symmetry of the classical theory may be destroyed in quantum 
theory by the quantum
anomaly and we have the corresponding anomalous WT identity. 
A well-known example
is the Adler-Bell-Jackiw(ABJ) anomaly. In the literature there are various 
ways to calculate
the anomaly, such as the point-splitting method \cite{JackiwJohnson}, 
the perturbative 
method \cite{ABJ} and Fujikawa's path integral method \cite{Fujikawa}. 
In a recent paper 
\cite{HeAnomaly} He 
calculated
the possible quantum anomaly of the transverse WT identities for the 
axial vector and vector
vertex in $U(1)$ gauge theories using the point-splitting method. 
The conclusion there is that
there is an anomaly(which is called transverse anomaly in that paper) 
in the axial vector 
current curl equation(and also the transverse WT identity for the axial 
vector vertex), while there
is no anomaly in the vector current curl equation. However, by careful 
inspections we find that the anomaly
term obtained in that paper actually equals zero. So the point-splitting 
method gives no transverse
anomaly for both the axial vector and vector current. The path integral 
treatment of these anomalies 
is also sketched by Kondo \cite{Kondo}. Due to the importance and subtleties of anomaly in field 
theory, it is necessary to reexamine this problem using various methods and see whether they give
the same result. In this paper we reexamine this problem using 
perturbative method and compare our result with that of He \cite{HeAnomaly}.  

First let us see the classical curl equations of the axial vector and vector current in a theory
of a classical Dirac field interacting with an external $U(1)$ gauge field. The Lagrangian is 
\begin{equation}
{\cal L}={\bar \psi}(x)(i\not\!D-m)\psi(x)
\end{equation}
where $D_\mu=\partial_\mu-
ieA_\mu(x)$ is the covariant derivative.
By using the equation of motion for the Dirac field, one can verify that
\begin{eqnarray}
&&\partial_\rho({\bar \psi}(x)\frac{i}{2}(\gamma^{\rho} \Omega-{\tilde \Omega}\gamma^{\rho})
\psi(x)) \nonumber \\
&=& {\bar \psi}(x)\frac{i}{2}(\gamma^{\rho} \Omega +{\tilde \Omega} \gamma^{\rho})(
\stackrel{\rightarrow}{\partial_\rho}-\stackrel{\leftarrow}{\partial_\rho})\psi(x)
+e{\bar \psi}(x)A_\rho(x)(\gamma^{\rho}\Omega+{\tilde \Omega}\gamma^{\rho})\psi(x)
\nonumber \\
&&-m{\bar \psi}(x)(\Omega+{\tilde \Omega})\psi(x)
\end{eqnarray}
Now if we choose $\Omega=-{\tilde \Omega}=\gamma_5 \sigma^{\mu\nu}$ in Eq.(3),
then by using the identity
$\{\gamma^{\rho},\gamma_5 \sigma^{\mu\nu}\}=2i(g^{\rho\mu} \gamma^{\nu} \gamma_5-g^{\rho\nu}
\gamma^{\mu}\gamma_5)$, we get the curl equation of the axial vector current:
\begin{eqnarray}
&&\partial^{\mu}({\bar \psi}(x)\gamma^{\nu}\gamma_5 \psi(x))-\partial^{\nu}({\bar \psi}(x)
\gamma^{\mu}\gamma_5\psi(x)) \nonumber \\
&=&-{\bar \psi}(x)\frac{i}{2}[\gamma^{\rho},\gamma_5 \sigma^{\mu\nu}](\stackrel{\rightarrow}
{\partial}_\rho-\stackrel{\leftarrow}{\partial}_\rho)\psi(x)-e{\bar \psi}(x)A_\rho(x)
[\gamma^{\rho},\gamma_5\sigma^{\mu\nu}]\psi(x)
\end{eqnarray}
Similarly, by choosing $\Omega={\tilde \Omega}=\sigma^{\mu\nu}$ and using the 
identity $[\gamma^{\rho},\sigma^{\mu\nu}]=2i(g^{\rho\mu}\gamma^{\nu}-g^{\rho\nu}\gamma
^{\mu})$, we get the curl equation of the vector current:
\begin{eqnarray}
&&\partial^{\mu}({\bar \psi}(x)\gamma^{\nu} \psi(x))-\partial^{\nu}({\bar \psi}(x)\gamma^
{\mu}\psi(x)) \nonumber \\
&=&2m {\bar \psi}(x) \sigma^{\mu\nu}\psi(x)-{\bar \psi}(x)\frac{i}{2}\{\gamma^{\rho},
\sigma^{\mu\nu}\}(\stackrel{\rightarrow}{\partial}_\rho-\stackrel{\leftarrow}{\partial
}_\rho)\psi(x)-e{\bar \psi}(x)A_\rho(x)\{\gamma^{\rho},\sigma^{\mu\nu}\}\psi(x)
\end{eqnarray}
Eqs.(4) and (5) are classical curl equations of the axial vector and vector current. Now let us 
quantize the Dirac field(but still treat the $U(1)$ gauge field as a classical external field).
Then the above curl equations may suffer from anomalies due to singularities of product of local 
operators. As we have mentioned, these possible "transverse anomalies" have been calculated in 
\cite{HeAnomaly} using the point-splitting method. In the following we shall calculate them using 
perturbative methods.

let us first consider the axial vector current. After quantization, the axial vector current 
curl equation may have an anomaly:
\begin{eqnarray}
&&\partial^{\mu}\langle {\bar \psi}(x)\gamma^{\nu}\gamma_5\psi(x) \rangle-\partial^{\nu}\langle
{\bar \psi}(x) \gamma^{\mu} \gamma_5 \psi(x) \rangle \nonumber \\
&=&-\langle {\bar \psi}(x)\frac{i}{2}[\gamma^{\rho},\gamma_5 \sigma^{\mu\nu}](\stackrel
{\rightarrow}{\partial_\rho}-\stackrel{\leftarrow}{\partial_\rho})\psi(x)\rangle -e\langle
{\bar \psi}(x)[\gamma^{\rho},\gamma_5 \sigma^{\mu\nu}]\psi(x)\rangle A_\rho(x)+anomaly
\end{eqnarray}
where $\langle \cdots \rangle$ stands for the vacuum expectation value in the presence of the 
external $U(1)$ gauge field $A_\mu(x)$. In the following we shall calculate perturbatively 
the vacuum expectation values appearing in Eq.(6) and see whether there exists an anomaly term.
The vacuum expectation value of each operator appearing in Eq.(6) is a functional of the 
external field 
$A_\mu$. After expanding it in powers of $A_\mu$ we get a series of one-loop diagrams. Now 
let us analyse these vacuum expectation values order by order in $A_\mu$.

The $A^n$ order contribution to $\langle {\bar \psi}(x)\gamma^{\nu} \gamma_5 \psi(x) \rangle$ is
represented by an $(n+1)$-gon diagram(see fig.1) and equals 
\begin{eqnarray}
\int \frac{d^4 q_1}{(2\pi)^4}\cdots \frac{d^4 q_n}{(2\pi)^4} e^{-i(q_1+\cdots +q_n)\cdot x}
(-1) \int\frac{d^4 k}{(2\pi)^4} tr[\gamma^{\nu}\gamma_5 \frac{i}{\not\! k-m}\gamma^{\rho_1}\frac
{i}{\not\! k-{\not\! q}_1-m}\cdots \nonumber \\
\times \gamma^{\rho_n}\frac{i}{\not\! k-{\not\! q}_1-\cdots-{\not\! q}_n-m}](ie)^n {\tilde A}_{\rho_1}
(q_1)\cdots {\tilde A}_{\rho_n}(q_n)
\end{eqnarray}
where ${\tilde A}_\rho(q)$ is the Fourier transform of $A_\rho(x)$:$A_\rho(x)=\int \frac{d^4 q}
{(2\pi)^4} e^{-iq\cdot x}{\tilde A}_\rho(q)$. Therefore the $A^n$ order contribution to  
$\partial^{\mu}\langle{\bar \psi}(x)\gamma^{\nu}\gamma_5\psi(x)\rangle-\partial^{\nu}
\langle{\bar \psi}(x)\gamma^{\mu}\gamma_5\psi(x)\rangle$ is given by 
\begin{eqnarray}
&&\int \frac{d^4 q_1}{(2\pi)^4}\cdots\frac{d^4 q_n}{(2\pi)^4} e^{-i(q_1+\cdots+q_n)\cdot x}\int\frac
{d^4 k}{(2\pi)^4} tr[i[(q_1+\cdots +q_n)^{\mu} \gamma^{\nu}-(q_1+\cdots +q_n)^{\nu} \gamma^{\mu}]
\gamma_5 \frac{i}{\not\! k-m}\gamma^{\rho_1} \nonumber \\
&&\times\frac{i}{\not\! k-{\not\! q}_1-m}\cdots \gamma^{\rho_n}\frac{i}{\not\! k-{\not\! q}_1-\cdots 
-{\not\! q}_n-m}](ie)^n {\tilde A}_{\rho_1}(q_1)\cdots{\tilde A}_{\rho_n}(q_n) \nonumber \\
&=&\int \frac{d^4 q_1}{(2\pi)^4}\cdots\frac{d^4 q_n}{(2\pi)^4} e^{-i(q_1+\cdots+q_n)\cdot x}\int\frac
{d^4 k}{(2\pi)^4} I_1^{\mu\nu\rho_1 \cdots \rho_n}(k,q_1\cdots q_n;m) 
(ie)^n {\tilde A}_{\rho_1}(q_1)\cdots{\tilde A}_{\rho_n}(q_n)
\end{eqnarray}
Because the axial vector current ${\bar \psi}(x)\gamma^{\mu}\gamma_5\psi(x)$ is even under charge 
conjugation, (8) is non-vanishing only for even
$n$. For low $n$, the loop integral in (8) is divergent and needs to be regularized. In this paper, 
the Pauli-Villars regularization is employed. More expicitly, we make the replacement
$I_1^{\mu\nu\rho_1 \cdots \rho_n}(k,q_1\cdots q_n;m)\rightarrow 
I_1^{\mu\nu\rho_1 \cdots \rho_n}(k,q_1\cdots q_n;m)+\sum_{s=1}^S C_s(m\rightarrow M_s)$, 
where $M_s$ are large masses which will eventually go to infinity and the coefficients $C_s$ 
are chosen to make the loop integral convergent. Now let us rationalize the denomenator, compute
the trace and expand the numerator and the denomenator in powers of $m^2$:
\begin{eqnarray}  
&&I_1^{\mu\nu\rho_1\cdots \rho_n}(k,q_1 \cdots q_n;m) \nonumber \\
&=& tr[i[(q_1+\cdots +q_n)^{\mu}\gamma^{\nu}-(q_1+\cdots +q_n)^{\nu}\gamma^{\mu}]\gamma_5
\frac{i}{\not\! k-m}\gamma^{\rho_1}\frac{i}{\not\! k-{\not\! q}_1-m}\cdots \gamma^{\rho_n}
\frac{i}{\not\! k-{\not\! q}_1-\cdots -{\not\! q}_n-m}] \nonumber \\
&=& \frac{Q_{n+1}+m^2Q_{n-1}+\cdots}{P_{2n+2}+m^2 P_{2n}+\cdots } \nonumber \\
&=& \frac{Q_{n+1}}{P_{2n+2}}+(\frac{Q_{n-1}}{P_{2n+2}}-\frac{P_{2n}Q_{n+1}}{P_{2n+2}^2})m^2+\cdots
\end{eqnarray}
where $P_m$ and $Q_m$ are polynomials in $k$ of degree less than or equal to $m$. 
The coefficient of $m^{2l}$ behaves as $\frac{1}{k^{n+1+2l}}$ for large $|k|$. So if we impose the two conditions
$1+\sum_{s=1}^SC_s=0$ and $m^2+\sum_{s=1}^S C_s M_s^2=0$,
the loop integral
\begin{equation}
\int \frac{d^4 k}{(2\pi)^4}(I_1^{\mu\nu\rho_1\cdots \rho_n}(k,q_1\cdots q_n;m)+\sum_{s=1}^S C_s (m\rightarrow M_s))
\end{equation}
will be convergent for all $n$. Such conditions may be realized through the introduction of only two auxiliary masses
$M_1^2=m^2+2\Lambda^2$ , $M_2^2=m^2+\Lambda^2$, where $\Lambda^2$ is a large cutoff which will ultimately
go to infinity.
This choice is such that $C_1=\frac{M_2^2-m^2}{M_1^2-M_2^2}=1$ and $C_2=\frac{M_1^2-m^2}{M_2^2-M_1^2}=-2$.
Now using $i[(q_1+\cdots + q_n)^{\mu}\gamma^{\nu}-(q_1+\cdots+ q_n)^{\nu}\gamma^{\mu}]=-\frac{1}{2}[\sigma^{\mu\nu},
{\not\! q}_1+\cdots +{\not\! q}_n]$ and ${\not\! q}_1+\cdots +{\not\! q}_n=(\not\! k-m)-(\not\! k-{\not\! q}_1-\cdots
-{\not\! q}_n-m)$, we can write 
\begin{eqnarray}
&&\int \frac{d^4 k}{(2\pi)^4}(I_1^{\mu\nu\rho_1\cdots \rho_n}(k,q_1\cdots q_n;m)+\sum_{s=1}^2 C_s(m\rightarrow M_s)) \nonumber \\
&=& \frac{1}{2}\int \frac{d^4 k}{(2\pi)^4}\{tr[\sigma^{\mu\nu}\gamma_5 i \gamma^{\rho_1}\frac{i}{\not\! k-{\not\! q}_1-m}\cdots
\gamma^{\rho_n}\frac{i}{\not\! k-{\not\! q}_1-\cdots -{\not\! q}_n-m} \nonumber \\
&&-\sigma^{\mu\nu}\gamma_5(\not\! k-{\not\! q}_1-\cdots - {\not\! q}_n-m)\frac{i}{\not\! k-m}\gamma^{\rho_1}
\frac{i}{\not\! k-{\not\! q}_1-m}\cdots \gamma^{\rho_n}\frac{i}{\not\! k-{\not\! q}_1-\cdots - {\not\! q}_n-m} \nonumber \\
&&+\sigma^{\mu\nu}\gamma_5 \frac{i}{\not\! k-m}\gamma^{\rho_1}
\frac{i}{\not\! k-{\not\! q}_1-m}\cdots \gamma^{\rho_n}\frac{i}{\not\! k-{\not\! q}_1-\cdots - {\not\! q}_n-m}(\not\! k-m)
\nonumber \\
&&-\sigma^{\mu\nu}\gamma_5\frac{i}{\not\! k-m}\gamma^{\rho_1}\frac{i}{\not\! k-{\not\! q}_1-m}\cdots \gamma^{\rho_n}i]
\nonumber \\
&&+\sum_{s=1}^2 C_s(m\rightarrow M_s)\}
\end{eqnarray}

The $A^n$ order contribution to $-\langle {\bar \psi}(x)
\frac{i}{2}[\gamma^{\rho},\gamma_5 \sigma^{\mu\nu}](\stackrel{\rightarrow}{\partial}_\rho-
\stackrel{\leftarrow}{\partial}_\rho)\psi(x)\rangle$ is represented by a similar $(n+1)$-gon diagram(see fig.1)
and equals
\begin{eqnarray}
&&\int \frac{d^4 q_1}{(2\pi)^4}\cdots \frac{d^4 q_n}{(2\pi)^4} e^{-i(q_1+\cdots+q_n)\cdot x}
\int \frac{d^4 k}{(2\pi)^4} tr[\frac{i}{2}[\gamma^{\rho},\gamma_5 \sigma^{\mu\nu}] \frac{i}
{\not\! k -m}
\gamma^{\rho_1}\frac{i}{\not\! k-{\not\! q}_1-m}
\cdots \gamma^{\rho_n} \nonumber \\
&&\times\frac{i}{\not\! k-{\not\! q}_1-\cdots -{\not\! q}_n-m}]((-i)k_\rho-i(k-q_1-\cdots-q_n)_\rho)(ie)^n
{\tilde A}_{\rho_1}(q_1)\cdots {\tilde A}_{\rho_n}(q_n) \nonumber \\
&=&\int \frac{d^4 q_1}{(2\pi)^4}\cdots \frac{d^4 q_n}{(2\pi)^4} e^{-i(q_1+\cdots+q_n)\cdot x}
\frac{1}{2}\int\frac{d^4 k}{(2\pi)^4}tr[[2\not\! k-{\not\! q}_1-\cdots -{\not\! q}_n, \gamma_5
\sigma^{\mu\nu}] \frac{i}
{\not\! k -m}
\gamma^{\rho_1} \nonumber \\
&&\times\frac{i}{\not\! k-{\not\! q}_1-m} 
\cdots \gamma^{\rho_n}\frac{i}{\not\! k-{\not\! q}_1-\cdots -{\not\! q}_n-m}](ie)^n
{\tilde A}_{\rho_1}(q_1)\cdots {\tilde A}_{\rho_n}(q_n) \nonumber \\
&=&\int \frac{d^4 q_1}{(2\pi)^4}\cdots \frac{d^4 q_n}{(2\pi)^4} e^{-i(q_1+\cdots+q_n)\cdot x}
\int\frac{d^4 k}{(2\pi)^4}I_2^{\mu\nu\rho_1\cdots\rho_n}(k,q_1\cdots q_n;m)(ie)^n
{\tilde A}_{\rho_1}(q_1)\cdots {\tilde A}_{\rho_n}(q_n) 
\end{eqnarray}
Because the operator ${\bar \psi}(x)\frac{i}{2}[\gamma^{\rho},\gamma_5\sigma^{\mu\nu}]
(\stackrel{\rightarrow}{\partial}_\rho-\stackrel{\leftarrow}{\partial}_\rho)\psi(x)$ is even under 
charge conjugation, (12) is non-vanishing only for even $n$. 
By similar arguments, one can show that the introduction of only two auxiliary masses
$M_1$ and $M_2$ can make the loop integral in (12) convergent( for $n \geq 1$).
Now using $2\not\! k-{\not\! q}_1-\cdots -{\not\! q}_n=(\not\! k-m)+(\not\! k
-{\not\! q}_1-\cdots -{\not\! q}_n-m)+2m$, we can write
\begin{eqnarray}
&& \int\frac{d^4 k}{(2\pi)^4}(I_2^{\mu\nu\rho_1\cdots \rho_n}(k,q_1\cdots q_n;m)+
\sum_{s=1}^2C_s(m\rightarrow M_s)) \nonumber \\
&=& \int\frac{d^4 k}{(2\pi)^4}\{\frac{1}{2}tr[\gamma_5 \sigma^{\mu\nu}\frac{i}{\not\! k-m}
\gamma^{\rho_1}\frac{i}{\not\! k-{\not\! q}_1-m}\cdots \gamma^{\rho_n}\frac{i}{\not\! k
-{\not\! q}_1-\cdots -{\not\! q}_n-m}(\not\! k-m) \nonumber \\
&&+\gamma_5\sigma^{\mu\nu}\frac{i}{\not\! k-m}\gamma^{\rho_1}\frac{i}{\not\! k-
{\not\! q}_1-m}\cdots \gamma^{\rho_n} i \nonumber \\
&&-\gamma^5\sigma^{\mu\nu}i\gamma^{\rho_1}\frac{i}{\not\! k-{\not\! q}_1-m}\cdots 
\gamma^{\rho_n}\frac{i}{\not\! k
-{\not\! q}_1-\cdots -{\not\! q}_n-m} \nonumber \\
&&-\gamma_5 \sigma^{\mu\nu}(\not\! k-{\not\! q}_1-\cdots -{\not\! q}_n-m)\frac{i}
{\not\! k-m}\gamma^{\rho_1}\frac{i}{\not\! k-{\not\! q}_1-m}\cdots 
\gamma^{\rho_n}\frac{i}{\not\! k
-{\not\! q}_1-\cdots -{\not\! q}_n-m}] \nonumber \\
&&+\sum_{s=1}^2 C_s(m\rightarrow M_s)\}
\end{eqnarray}

To calculate the $A^n$ order contribution to $-e\langle {\bar \psi}(x) [\gamma^{\rho},
\gamma_5\sigma^{\mu\nu}]\psi(x)\rangle A_\rho(x)$, one needs to calculate the $A^{n-1}$ 
order contribution to
$\langle {\bar \psi}(x)[\gamma^{\rho},\gamma_5\sigma^{\mu\nu}]\psi(x)\rangle$. The latter is represented by an
$n$-gon diagram(see fig.2) and equals 
\begin{eqnarray}
\int \frac{d^4 q_1}{(2\pi)^4}\cdots \frac{d^4 q_{n-1}}{(2\pi)^4}e^{-i(q_1+\cdots +q_{n-1})\cdot x}
(-1)\int \frac{d^4 k}{(2\pi)^4} tr[[\gamma^{\rho},\gamma_5 \sigma^{\mu\nu}]\frac{i}{\not\! k-m}
\gamma^{\rho_1}\frac{i}{\not\! k-{\not\! q}_1-m} \nonumber \\
\times\cdots\gamma^{\rho_{n-1}}\frac{i}{\not\! k-{\not\! q}_1-\cdots -{\not\! q}_{n-1}-m}](ie)^{n-1}
{\tilde A}_{\rho_1}(q_1)\cdots {\tilde A}_{\rho_{n-1}}(q_{n-1}) 
\end{eqnarray}
After multiplying by $e A_\rho (x) \rightarrow e A_{\rho_n}(x)=e\int \frac{d^4 q_n}{(2\pi)^4}
e^{-iq_n \cdot x}{\tilde A}_{\rho_n}(q_n)$, we get the $A^n$ order contribution to
$-e\langle {\bar \psi}(x)[\gamma^{\rho},
\gamma_5\sigma^{\mu\nu}]\psi(x)\rangle A_\rho(x)$:
\begin{eqnarray}
&&\int \frac{d^4 q_1}{(2\pi)^4}\cdots \frac{d^4 q_n}{(2\pi)^4} e^{-i(q_1+\cdots q_n)\cdot x}
(-i)\int \frac{d^4 k}{(2\pi)^4} tr[[\gamma^{\rho_n},\gamma_5 \sigma^{\mu\nu}]\frac{i}{\not\! k-m}
\gamma^{\rho_1} \frac{i}{\not\! k-{\not\! q}_1 -m} \nonumber \\
&&\times\cdots\gamma^{\rho_{n-1}}\frac{i}{\not\! k-{\not\! q}_1-\cdots -{\not\! q}_{n-1}-m}](ie)^n
{\tilde A}_{\rho_1}(q_1)\cdots {\tilde A}_{\rho_n}(q_n) \nonumber \\
&=&\int \frac{d^4 q_1}{(2\pi)^4}\cdots \frac{d^4 q_n}{(2\pi)^4} e^{-i(q_1+\cdots q_n)\cdot x}
\int\frac{d^4 k}{(2\pi)^4}I_3^{\mu\nu\rho_1\cdots \rho_n}(k,q_1\cdots q_n;m)(ie)^n
{\tilde A}_{\rho_1}(q_1)\cdots {\tilde A}_{\rho_n}(q_n) 
\end{eqnarray}
Because the operator ${\bar \psi}(x)[\gamma^{\rho},\gamma_5\sigma^{\mu\nu}]\psi(x)$ is odd under charge 
conjugation, (14)(and also (15)) is non-vanishing only for even $n$. 
The loop integral in (15) can also be regulated by introducing only two auxiliary masses 
$M_1$ and $M_2$. Now we write
\begin{eqnarray}
&&\int\frac{d^4 k}{(2\pi)^4}(I_3^{\mu\nu\rho_1\cdots \rho_n}(k,q_1\cdots q_n;m)+\sum_{s=1}^2
C_s(m\rightarrow M_s)) \nonumber \\
&=& \int\frac{d^4 k}{(2\pi)^4}\{(-i)tr[\gamma_5\sigma^{\mu\nu}\frac{i}{\not\! k-m}\gamma^{\rho_1}
\frac{i}{\not\! k-{\not\! q}_1-m}\cdots\gamma^{\rho_{n-1}}\frac{i}{\not\! k-{\not\! q}_1-\cdots -
{\not\! q}_{n-1}-m}\gamma^{\rho_n}] \nonumber \\
&&+\sum_{s=1}^2 C_s(m\rightarrow M_s)\} \nonumber \\
&&+\int \frac{d^4 k}{(2\pi)^4}\{i tr[\gamma_5\sigma^{\mu\nu}\gamma^{\rho_n}\frac{i}{\not\! k-m}\gamma^{\rho_1}
\frac{i}{\not\! k-{\not\! q}_1-m}\cdots\gamma^{\rho_{n-1}}\frac{i}{\not\! k-{\not\! q}_1-\cdots -
{\not\! q}_{n-1}-m}] \nonumber \\
&&+\sum_{s=1}^2 C_s(m\rightarrow M_s)\}
\end{eqnarray}
In the second term in the RHS of Eq.(16), we first rename the integration(summation) variables(indices):
$(q_n,\rho_n)\rightarrow (q_1,\rho_1),(q_1,\rho_1)\rightarrow(q_2,\rho_2), \cdots (q_{n-1},
\rho_{n-1})\rightarrow (q_n,\rho_n)$(this renaming does not alter its contribution to the 
integral $\int\frac{d^4 q_1}{(2\pi)^4}\cdots\frac{d^4 q_n}{(2\pi)^4}$ in (15)) and then make the shift  
$k \rightarrow k-q_1$ to obtain
\begin{eqnarray}
\int\frac{d^4 k}{(2\pi)^4}\{i tr[\gamma_5 \sigma^{\mu\nu}\gamma^{\rho_1}\frac{i}{\not\! k-{\not\! q}_1
-m}\gamma^{\rho_2}\frac{i}{\not\! k-{\not\! q}_1-{\not\! q}_2-m}\cdots \gamma^{\rho_n}
\frac{i}{\not\! k- {\not\! q}_1-{\not\! q}_2-\cdots -{\not\! q}_n-m}] \nonumber \\
+\sum_{s=1}^2 C_s (m\rightarrow M_s)\}
\end{eqnarray}
So (16) becomes 
\begin{eqnarray}
&&\int\frac{d^4 k}{(2\pi)^4}\{tr [-\gamma_5\sigma^{\mu\nu}\frac{i}{\not\! k-m}\gamma^{\rho_1}
\frac{i}{\not\! k-{\not\! q}_1-m}\cdots \gamma^{\rho_n}i \nonumber \\
&&+\gamma_5\sigma^{\mu\nu}i \gamma^{\rho_1}\frac{i}{\not\! k-{\not\! q}_1-m}\cdots \gamma^{\rho_n}
\frac{i}{\not\! k-{\not\! q}_1-\cdots-{\not\! q}_n-m}] \nonumber \\
&&+\sum_{s=1}^2 C_s (m\rightarrow M_s)\}
\end{eqnarray}

Comparing (11), (13) and (18), we find that the curl equation of the axial vector current is 
satisfied (at the level of vacuum expectation values) and that there is no transverse anomaly
for the axial vector current. At this point it is interesting to compare our result with that of \cite{HeAnomaly}
and see whether they agree with each other.. The anomalous axial vector current curl equation 
derived in \cite{HeAnomaly} reads
\begin{eqnarray}
&& \partial^{\mu}j_5^{\nu}(x)-\partial^{\nu}j_5^{\mu}(x) \nonumber \\
&=& \lim_{x' \rightarrow x} i(\partial_\lambda^{x}-\partial_\lambda^{x'})\varepsilon^{\lambda\mu\nu
\rho}{\bar \psi}(x')\gamma_\rho U_P(x',x)\psi(x) \nonumber \\
&&+\frac{g^2}{16\pi^2}[\varepsilon^{\alpha\beta\mu\rho}F_{\alpha\beta}(x)F^{\nu}_{~\rho}(x)
-\varepsilon^{\alpha\beta\nu\rho}F_{\alpha\beta}(x)F^{\mu}_{~\rho}(x)]
\end{eqnarray}
where $U_P(x',x)=e^{-ig\int_x^{x'} dy^{\rho}A_\rho(y)}$ is the Wilson line and the last term is the 
anomaly term. First let us observe that, by explicitly performing the differentiation operations, 
taking the limit
$x' \rightarrow x$ and using the identity$[\gamma^{\lambda},\gamma_5 \sigma^{\mu\nu}]=
-2\varepsilon^{\lambda\mu\nu\rho}\gamma_\rho$, we can put the first term in the RHS of Eq.(19) into the form
$-{\bar \psi}(x)\frac{i}{2}[\gamma^{\lambda},\gamma_5\sigma^{\mu\nu}](\stackrel{\rightarrow}{\partial}
_\lambda-\stackrel{\leftarrow}{\partial}_\lambda)\psi(x)+g{\bar \psi}(x)[\gamma^{\lambda},
\gamma_5\sigma^{\mu\nu}]\psi(x)A_\lambda(x)$, which agrees with the corresponding expression in the
axial vector curl equation in this paper(after identifying $g=-e$). Now let us see the anomaly term.
The anomaly term is proportional to $\varepsilon^{\alpha\beta\mu\rho}F_{\alpha\beta}(x)F^{\nu}_{~\rho}
(x)
-\varepsilon^{\alpha\beta\nu\rho}F_{\alpha\beta}(x)F^{\mu}_{~\rho}(x)$. When $\mu=\nu$, the above 
expression vanishes automatically. When $\mu \not= \nu$, one can also prove that $\sum_{\alpha \beta \rho}
\varepsilon^{\alpha\beta\mu\rho}F_{\alpha\beta}(x)F^{\nu}_{~\rho}(x)$ vanishes(here and in the 
following the summation convention is suspended) and therefore the anomaly term is in fact equal to zero. 
Let us use $\mu,\nu,
\sigma,\tau$ to denote the four distinct indices $0,1,2,3$. Due to the antisymmetry  of the 
Levi-Civita tensor and the field strength tensor, the summation index $\rho$ can only take on two
values $\sigma$ and $\tau$:
\begin{eqnarray}
\sum_{\alpha\beta\rho}\varepsilon^{\alpha\beta\mu\rho}F_{\alpha\beta}(x)F^{\nu}_{~\rho}(x)&=&
\sum_{\alpha\beta}\varepsilon^{\alpha\beta\mu\sigma}F_{\alpha\beta}(x)F^{\nu}_{~\sigma}(x)
+\sum_{\alpha\beta}\varepsilon^{\alpha\beta\mu\tau}F_{\alpha\beta}(x)F^{\nu}_{~\tau}(x) \nonumber \\
&=& 2\varepsilon^{\nu\tau\mu\sigma}F_{\nu\tau}(x)F^{\nu}_{~\sigma}(x)+2\varepsilon^{\nu\sigma
\mu\tau}F_{\nu\sigma}(x)F^{\nu}_{~\tau}(x) \nonumber \\
&=&2\varepsilon^{\nu\tau\mu\sigma}(F_{\nu\tau}(x)F^{\nu}_{~\sigma}(x)-F_{\nu\sigma}(x)F^{\nu}_{~\tau}
(x))
\end{eqnarray}
After expressing the field strengths in terms of potentials,  one finds that the 
above expression vanishes. So our perturbative calculation gives the same result with that in \cite{HeAnomaly}
using point-splitting method. 

By the parallel procedure one can find that the vector 
current curl 
equation is also satisfied at the level of vacuum expectation values and hence there is no transverse anomaly
for the vector current:
\begin{eqnarray}
&& \partial^{\mu}\langle{\bar \psi}(x)\gamma^{\nu}\psi(x)\rangle-\partial^{\nu}\langle
{\bar \psi}(x)\gamma^{\mu}\psi(x)\rangle \nonumber \\
&=& 2m \langle {\bar \psi}(x)\sigma^{\mu\nu}\psi(x)\rangle-\langle{\bar \psi}(x)\frac{i}{2}
\{\gamma^{\rho},\sigma^{\mu\nu}\}(\stackrel{\rightarrow}{\partial}_\rho-\stackrel{\leftarrow}
{\partial}_\rho)\psi(x)\rangle-e\langle{\bar \psi}(x)\{\gamma^{\rho},\sigma^{\mu\nu}
\}\psi(x)\rangle A_\rho(x)
\end{eqnarray}
This result is in agreement with that of \cite{HeAnomaly}.

At this point we remark that instead of Pauli-Villars regularization, one can also employ other regularization schemes
such as dimensional regularization
to calculate the transverse anomalies. Then there arises the problem of how to define $\gamma_5$ in $D$ dimensions.
One possible definition, due originally to 't Hooft and Veltman \cite{tHooft}, is to take $\gamma_5=i\gamma_0
\gamma_1\gamma_2\gamma_3$ in $D$ dimensions. This definition can correctly reproduce the ABJ anomaly.
Using this definition, one can easily verify that the whole arguments 
given above in the language of Pauli-Villars regularization can be presented in the language of dimensional 
regularization without essential changes, and the conclusions are the same.    

To summarize, in this paper we reexamine the problem of transverse axial vector and vector anomalies in four-dimensional
$U(1)$ gauge theories studied in \cite{HeAnomaly} using perturbative methods. We find that there are no transverse anomalies
for both the axial vector and vector current. We also find that the apparent transverse anomaly term for the axial vector
current obtained in \cite{HeAnomaly} in fact equals zero. Thus the absence of transverse anomalies for both the axial
vector and vector current is verified by at least two methods, the point-splitting method and the perturbative method. 
From the arguments given in this paper, one can also see that the Pauli-Villars regularization and dimensional regularization
gives the same result on the transverse anomaly of the axial vector and vector current. 

We thank Prof. Chao-hsi Chang, Prof. Xiao-fu L\"u 
and Prof. Hanxin He
for helpful discussions.
This work was supported in part by the National Natural Science Foundation of China
under Grant No.90103018,10175033 and 10135030.

\begin{figure}
\includegraphics{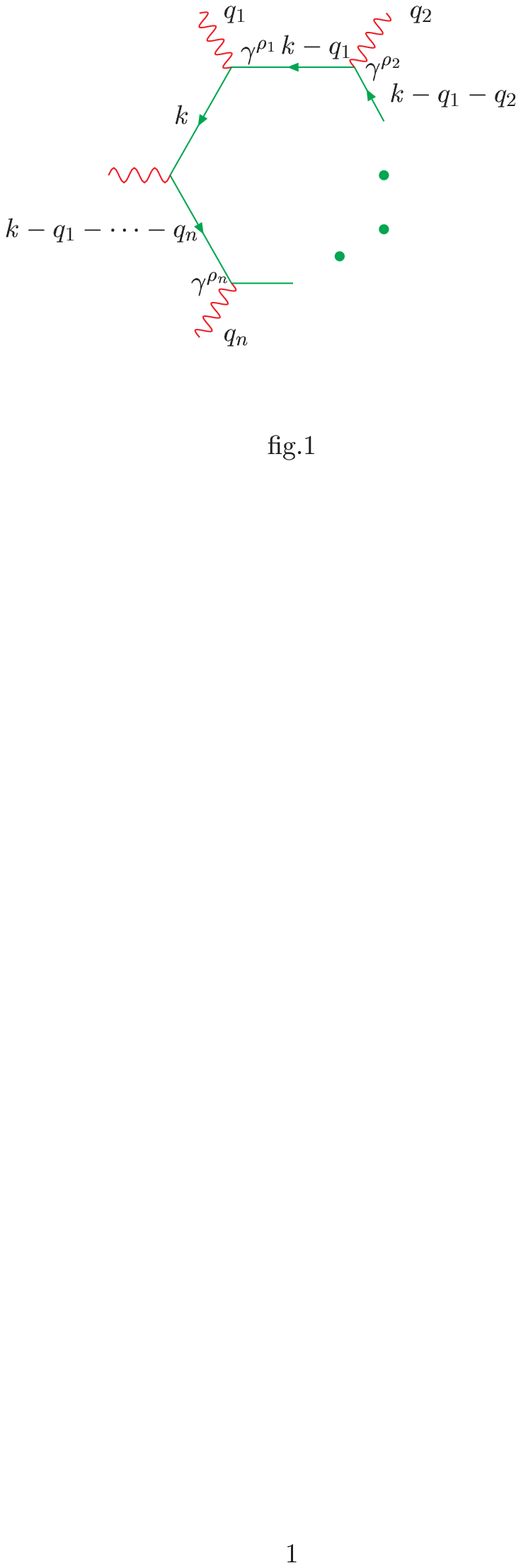}
\end{figure}
\begin{figure}
\includegraphics{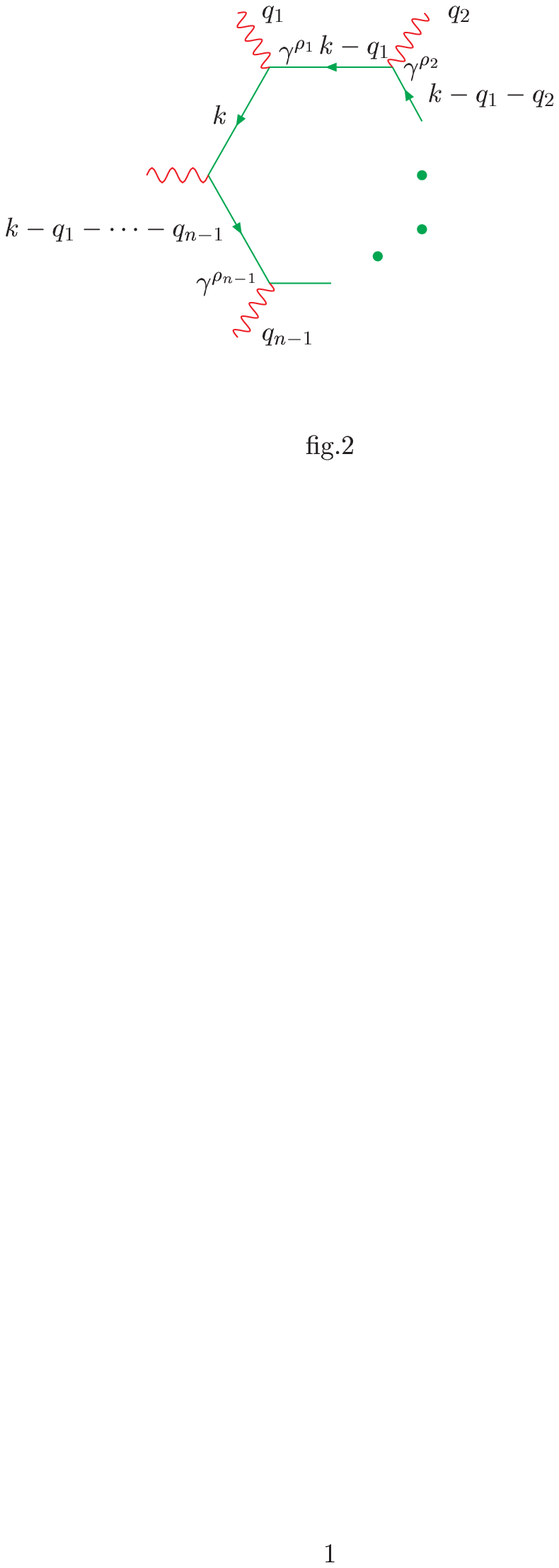}
\end{figure}

\end{document}